\documentstyle[12pt]{article}
\newcommand{\doublespace}{\renewcommand{\baselinestretch}{1.75}
   \Large\normalsize}

\renewcommand{\ref}[1]{\raisebox{.6ex}{[#1]}}

\newcommand{\be}{\begin{equation}}
\newcommand{\ee}{\end{equation}}

\newcommand{\ol}{\overline }

\begin{document}

\doublespace

\title{ Tunneling of a Quantized Vortex: Roles of
      Pinning and Dissipation }

\author{Ping Ao and David J. Thouless    \\
Department of Physics, FM-15                 \\
University of Washington, Seattle, WA 98195  \\ }

\maketitle

\begin{abstract}
We have performed a theoretical study of the effects of pinning
potential and dissipation on vortex tunneling in superconductors.
Analytical results are obtained in various limits relevant to experiment.
In general we have found that pinning and
dissipation tend to suppress the effect of the vortex velocity dependent part
of the Magnus force on vortex tunneling.
\end{abstract}

\noindent
PACS${\#}$s: 74.60.Ge; 05.40.+j; 30.70;-d; 74.20.-z

\newpage

To address the question of the resistance of a superconductor at zero
temperature, a clear understanding of vortex tunneling is needed.
Despite ample experimental evidence for vortex tunneling in
superconductors\cite{exp}, the theoretical study is of dimensional
analysis in nature\cite{blatter,fisher} and controversial\cite{ao1}.
Furthermore, in vortex tunneling
there is no comprehensive study of the role played by the Magnus force,
a force with a part proportional to the magnitude of vortex velocity
but perpendicular to the direction of the velocity.
This is a very unsatisfactory situation, in view of the fact that the
Magnus force is a general property of a vortex\cite{ao2} and tunneling is
extremely sensitive to dynamics.
In this paper we present an analytical study on vortex tunneling
in the presence of the Magnus force,
with a special attention paid to the effects of pinning and dissipation.
A general picture of those effects on vortex tunneling can be obtained from
our study.

We start with the Hamiltonian for a vortex.
To make the presentation simple, we consider vortex tunneling in a
semi-infinite superconductor film which defines the $x-y$ plane.
The supercurrent is along the $x$-direction, and the edge of the film
is lying on the $x$-axis. The length scale in the present study is always
much larger than the size of a vortex core.
A vortex can then be regarded as a point particle,
and as discussed in Ref.\cite{niu} its effective Hamiltonian
can be generally written as
\be
   H = \frac{1}{2M} [ {\bf P} - q_{v} {\bf A}({\bf r}) ]^{2} + V({\bf r})
   + \sum_{j} \left[ \frac{1}{2m_{j} } {\bf p}_{j}^{2} + \frac{1}{2} m_{j}
       \omega_{j}^{2} \left( {\bf q}_{j} - \frac{c_{j} }{m_{j}\omega_{j}^{2} }
                        {\bf r} \right)^{2} \right] \; ,
\ee
with the vector potential ${\bf A}$ determined by $
    \nabla\times {\bf A} = h \rho_{s} d \hat{z} /2$.
The meaning of each term in eq.(1) is as follows.
The vector potential reflects the existence of the the
vortex velocity dependent part (VVDP) of the Magnus force ${\bf F}_{m} = q_{v}
h \rho_{s} d ({\bf v}_{s} - \dot{\bf r} )\times \hat{z}/2$.
The Magnus force depends on the relative velocity  between the superfluid
velocity ${\bf v}_{s}$ and the vortex velocity $\dot{\bf r}$, and
the superfluid velocity dependent part (SFVDP) of the Magnus force will
contribute to the vortex potential $V$.
In accordance with the calculation of tunneling,
the vector potential will be taken as
\be
   {\bf A} =  \frac{ h }{2} \rho_{s} d \;  (y, 0, 0) \;,
\ee
and it can be shown that results are independent of the choice of gauge.
Here
$q_{v}= +1 (-1)$ stands for the vorticity parallel (anti-parallel) to the unit
vector $\hat{z}$ in $z$-direction, $h$ is the Planck constant, $\rho_{s}$ is
the superfluid electron number density with the factor 1/2 counting for the
Cooper pairing, and $d$ is the thickness of the film.
We take the vortex mass $M$ to be finite as demonstrated in Ref.\cite{niu},
and will show that it is relevant to vortex tunneling.
The vortex potential $V$ contains both the contributions from the SFVDP Magnus
force and pinning centers.
In the following we shall take $V$ to be of the form
\be
   V({\bf r}) = V_{1}(y) + \frac{1}{2} k_{x} x^{2} \; .
\ee
The pinning potential in the $x$-direction
is approximated by the harmonic potential,
and $k_{x}$ should be determined experimentally.
The potential $V_{1}(y)$ consists of the contributions from the SFVDP Magnus
force and the pinning potential in the $y$-direction, which
has a metastable point at $y=0$ in the present paper.
In the limit of no pinning and extreme type II,
\be
   V_{1}(y)= \frac{h}{2} \rho_{s} d \left( - v_{0} \; y +
                \frac{\hbar}{ 2m_{c} } \ln (y) \right),
\ee
only the contribution from the SFVDP Magnus: the potential due to
external supercurrent $v_{0}$
and the image potential from the edge of the superconductor film.
Here $m_{c}$ is the mass of a Cooper pair.
The dissipative environment of the vortex, the last term in eq.(1),
consists of a set of harmonic oscillators
as formulated in Ref.\cite{caldeira}.
The $j$th harmonic oscillator has a mass $m_{j}$ and a frequency $\omega_{j}$.
The effect of the dissipative
environment is specified by the spectral function
\be
   J(\omega) \equiv \pi \sum_{j} \frac{c_{j}^{2} }{2m_{j}\omega_{j} }
               \delta(\omega - \omega_{j} ) \; .
\ee
In the present paper, we shall assume the spectral function to have the
following form
\be
   J(\omega) = \eta \omega^{s}
         \exp\left\{ - \frac{\omega}{\omega_{c} } \right\} \; ,
\ee
with $\omega_{c}$ the cutoff frequency whenever needed.
In accordance with Ref.\cite{caldeira},
$s > 1 $ is the superohmic case, $s=1$ the ohmic case, and $0\leq s<1$
the subohmic case. In the ohmic damping case, $\eta$ is the vortex friction
such as discussed in Ref.\cite{bardeen}.

A remark on the relationship between the present study and others.
We note that the vortex motion is identical to the motion of an electron in the
presence of a magnetic field\cite{thouless}.
Results obtained in the present paper
can then be directly applied to that case with $q_{v}$ replaced by electron
charge $e$ and $h\rho_{s}/2$ by magnetic field $B$\cite{jain}.
For the case of a vortex in superfluid, the superfluid electron pair number
density $\rho_{s}/2$ should be
replaced by superfluid helium atom number density $\rho_{s}$\cite{muirhead}.

The tunneling is described by the Euclidean
action\cite{caldeira,thouless,jain}
\be
   S =     \int_{0}^{\hbar\beta } d\tau \left[ \frac{1}{2}
      M \dot{\bf r}^{2} + i \frac{h }{2} \rho_{s}d\; \dot{x} y +
       V_{1}(y) + \frac{1}{2} k_{x} x^{2}
    +  \sum_{j} \left( \frac{1}{2} m_{j} \dot{\bf q}_{j}^{2}
              + \frac{1}{2}m_{j}\omega_{j}^{2} \left({\bf q}_{j}
                           - \frac{c_{j} }{m_{j} \omega_{j}^{2} }
                       {\bf r} \right)^{2} \right) \right] \; ,
\ee
where $\beta = 1/k_{B}T$ is the inverse temperature.
The tunneling rate is equal to $\exp\{ - S_{c}/\hbar \} $, where
the semiclassical action $S_{c}$ is determined by the
bounce solution of the equation $\delta S =0$, in which the periodic boundary
condition, $({\bf r}(\hbar\beta), \{q_{j}(\hbar\beta)\}) = ({\bf r}(0),
\{q_{j}(0)\} )$, is required.

We are interested in the vortex tunneling out of the metastable state $y=0$.
After the
tunneling in $y$-direction other degrees of freedoms can take arbitrary value.
Therefore the summation over the final states,
integrations over environmental degrees of freedoms, $\{ {\bf q}_{j} \} $,
and over the $x$ coordinate, will be taken. Those are gaussian integrals.
After the integrations, the effective Euclidean action is
\be
    S_{eff} = \int_{0}^{\hbar\beta } d\tau
    \left[ \frac{1}{2} M \dot{y}^{2} + V_{1}(y) \right] +
     \frac{1}{2} \int_{0}^{\hbar\beta } d\tau \int_{0}^{\hbar\beta } d\tau'
     \left[ k(|\tau - \tau' |)  + g(\tau - \tau' ) \right]
     [ y(\tau) - y(\tau') ]^{2}  \; ,
\ee
with the normal damping kernel $k$ as
\be
   k(\tau) = \frac{1}{\pi} \int_{0}^{\infty} d\omega \;  J(\omega)
             \frac{ \cosh[\omega (\frac{\hbar\beta}{2} - \tau ) ] }
                  { \sinh[\frac{\omega\hbar\beta}{2} ] } \; ,
\ee
and the damping kernel due to the VVDP Magnus force  coupled $x$-direction
motion, which we shall call the anomalous damping kernel, as
\be
   g(\tau) =
      \frac{1}{\hbar\beta}\frac{1}{2M} \left( \frac{h}{2}\rho_{s}d \right)^{2}
             \sum_{n=-\infty}^{\infty}
   \frac{M \omega_{x}^{2} + \xi_{n} }
        {M \nu_{n}^{2} + M \omega_{x}^{2} + \xi_{n} } e^{ i \nu_{n}\tau } \; .
\ee
Here
\be
   \xi_{n} = \frac{1}{\pi} \int_{0}^{\infty} d\omega
             \frac{J(\omega ) }{\omega}
             \frac{2\nu_{n}^{2} }{\omega^{2} + \nu_{n}^{2} } \; ,
\ee
\be
   \nu_{n} = \frac{2\pi}{\hbar\beta} n \; ,
\ee
and
\be
   \omega_{x}^{2} = \frac{k_{x} }{M} \; .
\ee
In the large $\tau$ limit, from eqs.(6) and (9) the normal damping kernel
takes the form
\be
   k(\tau) = \frac{1}{\pi} \; \eta \; \frac{1}{\tau^{s+1} } \; ,
\ee
which demonstrates the one-to-one correspondence between the low frequence part
of the spectral function and the long time behavior of the damping kernel.
Now we have obtained an effective one-dimensional problem. This is the
starting point for the following detailed study of the tunneling in
various experimentally accessible limits.

{\it  no pinning, no dissipation } {\ }
Without disorder and dissipation, there is no quantum decay of
supercurrent in an infinite film\cite{ao4}.
In reality such as discussed in the present paper
the film always has an edge, tunneling is possible.
To see the role played by the VVDP Magnus force, we first
study the case with no dissipation and no pinning.
Because of the VVDP Magnus force,
motions in the $x$- and $y$- directions become coupled to each other.
Therefore although the normal damping kernel $k$ vanishes now, the $n=0$ mode
in the anomalous kernel $g$ contributes to the effective action.
Then the effective action is
\be
   S_{eff} = \int_{0}^{\hbar\beta } d\tau \left[ \frac{1}{2}
      M \dot{y}^{2} + V_{1}(y)  \right]
    + \frac{1}{4M} \left(\frac{h}{2} \rho_{s}d \right)^{2} \frac{1}{\hbar\beta}
      \int_{0}^{\hbar\beta} d\tau \int_{0}^{\hbar\beta} d\tau'
      [ y(\tau) - y(\tau') ]^{2} \; .
\ee
We note that in the language of spectral function, this corresponds to the
subohmic bath case with $s=0$({\it c.f.} eqs.(8) and (14)), and
eq.(15) is explicitly gauge invariant under the change $y \rightarrow
y + \; constant$, because of the periodic boundary condition of $x$ imposed in
the tunneling calculation.

For low enough density $\rho_{s}$, that is, for a weak VVDP Magnus force,
there is a tunneling solution. Particularly,
for very low density and low temperatures
the semiclassical action may be evaluated perturbatively:
\be
   S_{c} = \int_{-\infty}^{\infty } d\tau \left[ \frac{1}{2}
      M \dot{y}^{2}_{c}(\tau) + V_{1}( y_{c}(\tau) )  \right]
    + \frac{1}{2M} \left( \frac{h}{2} \rho_{s}d \right)^{2}
      \left[ \int_{-\infty}^{\infty} d\tau  y_{c}^{2}(\tau)
    - \frac{1}{\hbar\beta} \left( \int_{-\infty}^{\infty}
                                 d\tau  y_{c}(\tau) \right)^{2} \right] \; ,
\ee
where $y_{c}(\tau)$ is the bounce solution at zero temperature without
the VVDP Magnus force.
Eq.(16) shows a remarkable $\rho_{s}^{2}$ dependence and
linear temperature dependence\cite{ao3}.

For high enough density, that is, for a strong VVDP Magnus force,
the tunneling rate vanishes at zero temperature, because the VVDP Magnus force
renormalizes the original potential such that the state near $y=0$ is stable.
This can be noted from eq.(15) where as $T \rightarrow 0$
the cross term in the last term disappears.
A straightforward calculation leads to the criterion for the
localization as
\be
    \frac{1}{M} \left( \frac{h}{2}\rho_{s}d \right)^{2} >
     \left|\frac{d^{2} V_{1}(y) }{dy^{2} } \right|_{barrier\; top} \; .
\ee
Thus we have obtained that the $s=0$ dissipative environment is marginal
for localization in tunneling decay, compared to the $s=1$ case for the
tunneling splitting\cite{leggett}.
This is the dynamical localization caused by the VVDP Magnus force\cite{ao4}.
It is worthwhile to point out that according to eq.(17)
although a large VVDP Magnus force inhibits
vortex tunneling, a large vortex mass instead favors vortex tunneling.

{\it no pinning, finite dissipation} {\ }
Now we set the pinning potential $k_{x}/M = \omega^{2}_{x} = 0$ in eq.(10).
Both the normal damping kernel and the anomalous damping kernel are finite.
Since the quantum tunneling corresponds
to the limit of large imaginary time, $\hbar\beta \rightarrow \infty$,
we look for the large time limit behavior of the anomalous damping kernel $g$.
In the this limit, we may replace
the summation $1/\hbar\beta \sum_{n}$ by the integration $ 1/2\pi \int d\nu $.
Then for the environment $0 < s < 2$,
we find the anomalous damping kernel $g$ in the large $\tau$ limit as
\be
   g(\tau) =  a \; \frac{1}{2\pi M}
                \left( \frac{h}{2} \rho_{s}d \right)^{2} \;
          \frac{ 1 }{ \ol{\eta} } \; \frac{1}{\tau^{2-s+1} } \; ,
\ee
with $a$ a numerical constant of order of unity.
Here
\be
   \ol{\eta} = \frac{2}{\pi} \frac{\eta}{M} \int_{0}^{\infty} dz
                     \frac{z^{s-1} }{z^{2} + 1 } \; ,
\ee
and $J(\omega) = \eta \omega^{s} $ has been used.
The effective dissipative environment corresponding to the anomalous
damping kernel is $s_{eff} = 2-s$({\it c.f.} eqs.(6) and (14)),
which leaves the ohmic damping unchanged, transforms the subohmic damping
into superohmic damping and {\it vice versa}.

For the case $s> 2$, using eq.(4) for the spectral function $J$,
we find that the effective dissipative environment corresponding
to the anomalous damping kernel is
$s_{eff} = 0$, which smoothly connects the result for $0 < s < 2$.

An important example is the ohmic damping case, where an exact
expression for $a$ in eq.(18) can be obtained.
Carrying out a detailed but straightforward calculation,
we find the effect action at zero temperature as
\be
   S_{eff} = \int_{-\infty}^{\infty} d\tau
    \left[ \frac{1}{2} m \dot{y}^{2} + V_{1}(y) \right]
        + \frac{1}{2\pi} \; \eta_{eff}
         \int^{\infty}_{-\infty} d\tau \int^{\infty}_{-\infty}d\tau'
         \frac{1}{|\tau-\tau' |^{2} } [ y(\tau) - y(\tau) ]^{2} \; ,
\ee
with the effective damping strength $\eta_{eff}$ as
\be
   \eta_{eff} = \eta + \left( \frac{h\rho_{s} d }{2} \right)^{2}
                \frac{1}{\eta} \; .
\ee

The effect of a dissipative environment can now be summarized as follows.
It has been demonstrated in Ref.\cite{leggett} that
the subohmic dissipation has strong effects
on tunneling, while the superohmic dissipation has weak effects.
{}From the above analysis
we have that if the normal damping kernel $k(\tau)$ of eq.(9) is subohmic,
the anomalous damping kernel $g(\tau)$ of eq.(10) is superohmic, and {\it vice
versa}.
Then according to Ref.\cite{leggett} the effect of the VVDP Magnus force,
represented by the anomalous damping kernel, is weak/strong
on vortex tunneling for the normal subohmic/superohmic damping.
In particular, for normal superohmic damping with $s > 2$,
the vortex tunnels as if there were no effect of dissipation.
For the normal ohmic damping with $s=1$,
we need to compare the relative strength of the VVDP Magnus force and the
dissipation according to eq.(21).
Classically, the VVDP Magnus force tends to keep a vortex moving along
an equal potential contour, but the friction instead
along the gradient of the potential.
In general, we can conclude that the dissipation
tends to suppress the effect of the VVDP Magnus force on vortex tunneling.

{\it finite pinning, no dissipation} {\ }
The normal damping kernel vanishes in this situation. The anomalous damping
kernel can be expressed by hyperbolic functions.
We find that the effective action is then
\[
   S_{eff} = \int^{\hbar\beta}_{0} d\tau
             \left[ \frac{1}{2} M \dot{y}^{2}+V_{1}(y) \right]
   + \frac{1}{4M} \left( \frac{h}{2} \rho_{s}d \right)^{2}
     \int^{\hbar\beta}_{0} d\tau  \int^{\hbar\beta}_{0} d\tau'  \times
\]
\be
    \frac{\omega_{x}}{2}
    \frac{\cosh[\omega_{x} ( \frac{\hbar\beta}{2} - |\tau-\tau'|)] }
         {\sinh[\frac{\omega_{x}\hbar\beta}{2} ] }
    [ y(\tau) - y(\tau') ]^{2} \; .
\ee
It is the superohmic case with $s_{eff} = \infty$,
because the effective spectral function has the form
$\delta(\omega -\omega_{x} )$, which has no low frequency mode.
Therefore the tunneling rate is nonzero for any magnitude of the VVDP Magnus
force.
This result shows that pinning has a very strong influence on
the vortex tunneling in the presence of the VVDP Magnus force,
because the introducing of the pinning potential bends the straight line
trajectory of a vortex and makes the transition to other trajectories possible.
We note that by letting $\omega_{x} =0$ we recover eq.(15), where
the tunneling rate vanishes for a sufficiently strong VVDP Magnus force.

We can evaluate the semiclassical action perturbatively,
if the density is low, or, pinning potential is strong as done in eq.(16).
In the large density limit semiclassical action may be evaluated by a
variational method similar to Ref.\cite{caldeira} in the case of ohmic damping.
However, we have a more powerful method to perform the calculation,
because the strong VVDP Magnus force freezes the kinetic energy of the vortex.
In this case $x$ and $y$ coordinates now form a pair of
canonically conjugate variables,
and the Euclidean action is\cite{thouless,jain}
\be
   S_{eff} = \int^{\hbar\beta}_{0} d \tau
             \left[ i \frac{h}{2}\rho_{s}d \dot{x} y +
       V_{1}(y) + \frac{1}{2}k_{x} x^{2} \right] \; .
\ee
Following the calculation outlined in Refs.\cite{thouless,jain},
we find the semiclassical action in the form
\be
   S_{c} = h \; \rho_{s}d \; \int^{y_{t}}_{0 } dy
           \sqrt{  \frac{ 2 V_{1}(y) }{k_{x} }  } \; ,
\ee
with $y_{t}$ the turning point determined by the equation $V_{1}(y) =
0$\cite{feigelman}.
The result shows that as $k_{x}$ increases, the
semiclassical action decreases.
Therefore we conclude that the pinning potential in the $x$-direction
helps the tunneling in $y$-direction\cite{thouless}.

{\it general case: finite pinning and dissipation } {\ }
In general we need to go back to eqs.(8-10) to study the tunneling under the
influence of pinning and dissipation.
However, based on the insight gained by the above analysis
here we can draw a general conclusion without a detailed study:
Since both the dissipation and pinning tend to suppress the effect
of the VVDP Magnus force, the total effect of them will do the same.
Indeed the anomalous damping kernel is always superohmic.
In particular, in the presence of strong pinning and ohmic damping, the
superohmic-like anomalous damping kernel $g$
may be ignored compared to the ohmic normal damping kernel $k$.
Then from eq.(8) we have the effective action as
\be
    S_{eff} = \int_{0}^{\hbar\beta } d\tau
    \left[ \frac{1}{2} M \dot{y}^{2} + V_{1}(y) \right] +
     \frac{\eta}{2\pi} \int_{0}^{\hbar\beta }
      d\tau \int_{0}^{\hbar\beta } d\tau'
      \frac{1}{ |\tau - \tau' |^{2} }  [ y(\tau) - y(\tau') ]^{2}  \; ,
\ee
which looks as if there were no effect of the VVDP Magnus force.
This may explain the pronounced experimental observation of the  absence of
the effect of the VVDP Magnus force in vortex tunneling experiments
on dirty superconductors\cite{exp}.

To summarize, we have performed a complete analytical study of the influence
of pinning and dissipation on vortex tunneling.
The VVDP Magnus force tends to decrease the tunneling rate.
In the absence of pinning and dissipation,
there is no tunneling for a strong VVDP Magnus force.
Detailed results have been obtained when one or all of
them is absent.
Both pinning and dissipation tend to suppress the influence of the
VVDP Magnus force on vortex tunneling.
Present results may explain the absence of effects of the VVDP Magnus
force in vortex tunneling experiments on dirty superconductors.

\noindent
{\bf Acknowledgements:}
We thank Moo Young Choi, Qian Niu, and Yong Tan
for many helpful discussions.
This work was supported by US National Science Foundation under Grant No's.
DMR-8916052 and DMR-9220733.


\begin{thebibliography}{99}
\bibitem{exp}
 A.V. Mitin, Zh. Eksp. Teor. Fiz. {\bf 93}, 590 (1987);
 A.C. Mota, P. Visani, and A. Pollini. Phys. Rev. {\bf B37}, 9830 (1988);
 N. Giordano, Phys. Rev. Lett. {\bf 61}, 2137 (1988);
 A.C. Mota, G. Juri, P. Visani, A. Pollini, T. Teruzzi, K. Aupke, and B.
      Hilti, Physica {\bf C185-189}, 343 (1991);
 L. Fruchter, A.P. Malozemoff, I.A. Campbell, J. Sanchez, M. Konczykowski, R.
      Griessen, and F. Holtzberg, Phys. Rev. {\bf B43}, 8709 (1991);
 Y. Liu, D.B. Haviland, L. Glazman, and A.M. Goldman,
                Phys. Rev. Lett. {\bf 68}   (1992);
 D. Prost, L. Fruchter, I.A. Campbell, N. Motohira, and M. Konczykowski,
                Phys. Rev. {\bf B47}, 3457 (1993);
 J. Tejada, E.M. Chundnovsky, and A. Garcia, {\it ibid}, 11552 (1993).
\bibitem{blatter}
 G. Blatter, V.B. Geshkenbein, and V.M. Vinokur, Phys. Rev. Lett. {\bf 66},
     3297 (1991);
 G. Blatter, V.B. Geshkenbein, Phys. Rev. {\bf B47}, 2725 (1993).
\bibitem{fisher}
 M.P.A. Fisher, T.A. Tokuyasu,  and A.P. Young, Phys. Rev. Lett. {\bf 66},
        2931 (1991).
\bibitem{ao1}
 P. Ao, Phys. Rev. Lett. {\bf 69}, 2997 (1992).
\bibitem{ao2}
 P. Ao and D.J. Thouless, Phys. Rev. Lett. {\bf 70}, 2158 (1993);
 P. Ao, Q. Niu, and D.J. Thouless, Physica {\bf B\& C}, in print.
\bibitem{niu}
 Q. Niu, P. Ao, and D.J. Thouless, submitted to Phys. Rev. Lett.
\bibitem{caldeira}
  A.O. Caldeira and A.J. Leggett, Ann. Phys.(NY), {\bf 149}, 374 (1983);
    {\bf 153}, 445(E) (1984).
\bibitem{bardeen}
 J. Bardeen and M.J. Stephen, Phys. Rev. {\bf 140}, A1197 (1965).
\bibitem{thouless}
 D.J. Thouless, P. Ao, and Q. Niu, Physica {\bf A}, in print.
\bibitem{jain}
 J.K. Jain and S. Kivelson, Phys. Rev. {\bf A36}, 3467 (1987) and
              Phys. Rev. {\bf B37}, 4111 (1988);
 H.A. Fertig and B.I. Halperin, Phys. Rev. {\bf B36}, 7969 (1987).
\bibitem{muirhead}
 C.M. Muirhead, W.F. Vinen, and R.J. Donnelly, Phil. Trans. R. Soc. Lond.
            {\bf A311}, 433 (1984).
\bibitem{ao4}
 P. Ao, J. Low Temp. Phys. {\bf 89}, 543 (1992).
\bibitem{ao3}
 Same result was obtained for the electron tunneling in the presence of a
 magnetic field by P. Ao[Mod. Phys. Lett. {\bf B}, in print.]
\bibitem{leggett}
 A.J. Leggett, S. Chakravarty, A.T. Dorsey, M.P.A Fisher, A. Garg, W. Zwerger,
 Rev. Mod. Phys. {\bf 59}, 1 (1987).
\bibitem{feigelman}
 A special case of eq.(24) was obtained by
 M.V. Feigel'man, V.B. Geshkenbein, A.I. Larkin, and S. Levit[Preprint].

\end{thebibliography}
\end{document}